# A 4-state solution to the Firing Squad Synchronization Problem based on hybrid rule 60 and 102 cellular automata

LI Ning[1], LIANG Shi-li[1*], CUI Shuang[1], XU Mei-ling[1], ZHANG Ling[2]

(1. Department of Physics, Northeast Normal University, Changchun, 130024, China; 2. College of science, ChangChun University of Science and Technology, ChangChun, 130022,China)

**Abstract:**

In this paper, we present a 4-state solution to the Firing Squad Synchronization Problem (FSSP) based on hybrid rule 60/102 Cellular Automata(CA) ,This solution solves the problem on the line of length $2^n$ with two generals. Previous work on FSSP for 4-state systems focused mostly on linear cellular automata, where synchronizes an infinite number of lines but not all possible lines. We give time-optimal solutions to synchronize an infinite number of lines by rule 60 and rule 102 respectively, and construct a hybrid rule 60 and 102 states transition table. Compared to the known solutions of cellular automata, the hybrid CA way is simpler and faster, the minimal time is (n-1) step.

1. Introduction

The Firing Squad Synchronization Problem (FSSP), is one of the best studied problems for cellular automata. This problem in which a network of identical cells(finite automata) work synchronously at discrete time steps. Figure 1 shows a finite one-dimensional cellular array consisting of n cells, All cells (soldiers) except the left end cell (general), are initially in the quiescent state at time t=0. The next state is determined by both its own state and the present states of its right and left neighbors. At some future time, all of the cells will simultaneously and, for the first time, enter a special firing state.

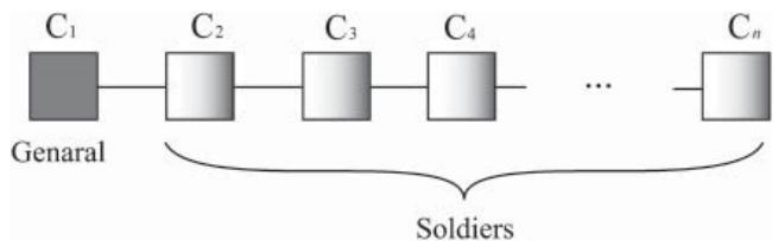

[*]Corresponding author. E-mail: lsl@nenu.edu.cn, Tel.: +86-431-85099665

Fig. 1.  A one-dimensional cellular automaton.

The FSSP problem has been studied for a long time ([1]). Minsky and McCarthy ([2])constructed a solution synchronizing n cells in 3n steps using a divide-and-conquer method. After that Waksman ([3]) and Balzer ([4]) designed a new algorithm which can synchronize in minimal time in T (n) =(2n−2)steps, and have a state complexity of 16 and 8 states respectively. Mazoyer ([5]) showed a 6-state minimal-time solution，but still unknown whether a five-state full solution exists，Yunès ([6, 7]), Settle and Simon ( [8]) and Umeo ( [9]) designed non minimal-time solutions with few states. Umeo ([10,11]) can find less than 6-state solutions to synchronize an infinite number. When it comes to four-state special solution, some people try to find solutions to synchronize an infinite number but not all lines. Balzer proved that the minimal time 4-state solution does not exist, and that Balzer's result shows that there is no 4-state solution which synchronizes every line in minimal-time. Now the 4-state solutions are all built using some elementary algebra, and mostly based on linear cellular automata with rule 60 and 150, Jean-Baptiste ([12]) presented one of these solutions in only 4-state which synchronizes every line whose length is a power of 2. In this paper, we quest a 4-state solution to a line of $2^n$ FSSP with two generals based on hybrid rule 60/102 Cellular Automata.

2.  Preliminary

For the four-state FSSP solution, Yunès designed an algorithm by Wolfram Rule 60 for $2^n$ cell array to meet an infinite number of lines. Our construction is also based on Wolfram's linear cellular automata using rule 60 and rule 102.Here are some basic FSSP definitions and elementary concepts.

2.1 Definitions

There is a cellular automata A（Q, $\delta$）  where Q is a finite set, called the states set of A, and $\delta$ is a transition function from $Q^3$ ->Q.

a)  A cellular automata S is a couple A which is an application from Z in Q. A configuration C evolves to another configuration C* so that
$$C^*（Z）= \delta（C(z-1), C(z), C(z+1)）$$

We can define C*= $\Delta$（C）as global transition function.
So the initial configuration of cellular automata is $C_0$ (at the time of t=0), the configuration of time t is $C_t = \Delta^t（C）$.

b)  At least four distinguished states belong to Q.

c)  State Q is the quiescent state. It satisfies $\delta$（Q, Q, Q）=Q, $\delta$（Q, Q, !）=Q, $\delta$（!, Q, Q）=Q

d)  State * is the boundary state. It satisfies: $\forall q_1, q_2 \in Q, \delta(q_1, *, q_2) = *$

e)  State G is the general state and the state F is the Fire state, such that, starting from the

initial configuration defined by:

$$(a) \forall z \leq 0, C[n](z) = !$$
$$(b) \forall z \geq n+1, C[n](z) = !$$
$$(c) C[n](1) = G$$
$$(d) \forall z \in \{2, 3, \ldots, n\}, C[n](z) = Q$$

f) The evolution of the configuration C[n] is such that, for synchronization time t(n):

$$(a) \forall z \in N, \forall t \in \{1, \ldots, t(n)-1\}, C[n]_t(z) \neq F$$
$$(b) \forall z \in \{1, \ldots, n\}, C[n]_{t(n)}(z) = F$$

2.2 Some FSSP elementary concepts

a) Synchronization Time[3] [4] [9] [10] [13].

The solution to the firing squad synchronization problem can be shown: Synchronization of n cells in less than 2n-2 steps is impossible, and Synchronization of n cells in exactly 2n-2 steps is possible. So the minimal synchronization time is 2n-2 steps.

b) The number of states[4] [9] [10].

To design a transition tables, there is at least three states: the state of quiescent, the state of general and the state of firing. Besides, we also design a boundary state. It is shown that there exists no three-state solution and no four-state symmetric solution on rings. Balzer proved that there is no four-state full solution for this problem. A minimal –time solution of six states was introduced by Jacques Mazoyer in 1987[13].

c) The number of transition rules[9] [10].

Any k-state transition table for the synchronization has at most $(k-1)k^2$ entries in (k-1) matrices of size k*k. The number of transition rules reflects the complexity of synchronization algorithms.

3. Yunès's Rule 60 algorithm

Yunès[12] designed four states solution to achieve this synchronization, the state general (g) is $\begin{bmatrix}1\\0\end{bmatrix}$, the quiescent state • is $\begin{bmatrix}0\\0\end{bmatrix}$, the state $x_1$ is $\begin{bmatrix}0\\1\end{bmatrix}$, and the state of firing is $\begin{bmatrix}1\\1\end{bmatrix}$, the symbol $ represents the boundary states. The rule 60 evolution figure is fig 2:

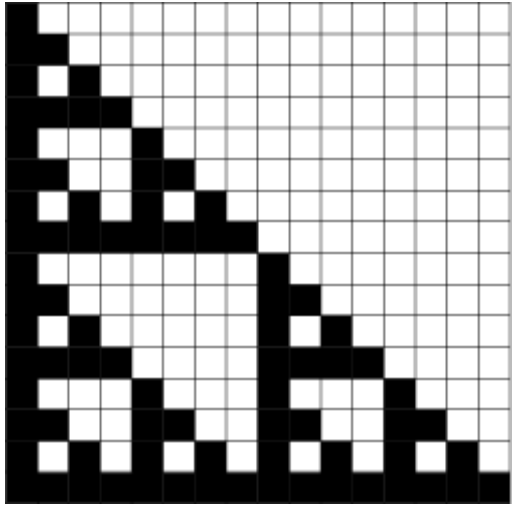

Fig. 2. The Rule 60 Cellular Auotmata

If we fold the pascal's triangle modulo2, we can get follow figure 3:

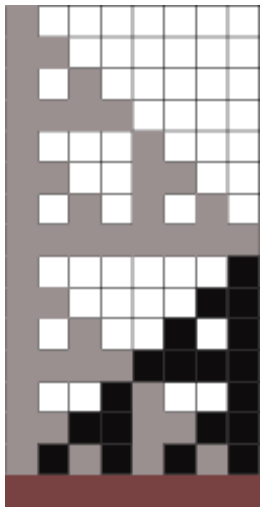

Fig. 3. The Rule 60 based solution.

The gray color states represent the state of general, the white color states represent the state of quiescent, the black color states represent the state of $x_1$, and the dark red color states represent the state of firing. Meanwhile, Yunès rewrite the transition function into algebraic form:

$$\begin{cases} \delta\left(\begin{bmatrix}a\\b\end{bmatrix}, \begin{bmatrix}c\\d\end{bmatrix}, \begin{bmatrix}e\\f\end{bmatrix}\right) = \begin{bmatrix}a+c(\mathrm{mod}\ 2)\\d+f(\mathrm{mod}\ 2)\end{bmatrix} \\ \delta\left(\$, \begin{bmatrix}c\\d\end{bmatrix}, \begin{bmatrix}e\\f\end{bmatrix}\right) = \begin{bmatrix}c\\d+f(\mathrm{mod}\ 2)\end{bmatrix} \\ \delta\left(\begin{bmatrix}a\\b\end{bmatrix}, \begin{bmatrix}c\\d\end{bmatrix}, \$\right) = \begin{bmatrix}a+c(\mathrm{mod}\ 2)\\c\end{bmatrix} \end{cases}$$

4. Rule 102 algorithm

    Based on the work of Yunès, we design a four-state solution by rule 102, in this way the general is located in the right end of cell array. The rule 102 evolution process can be seen in fig. 4:

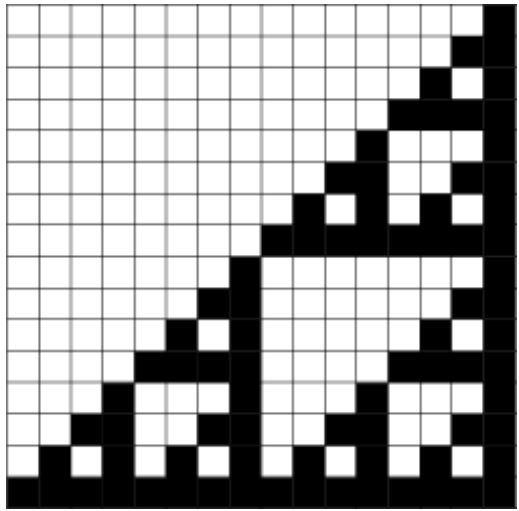

Fig.4. the Rule 102 Cellular Automata n=16

If we fold the pascal's triangle of rule 102, we will get fig. 5:

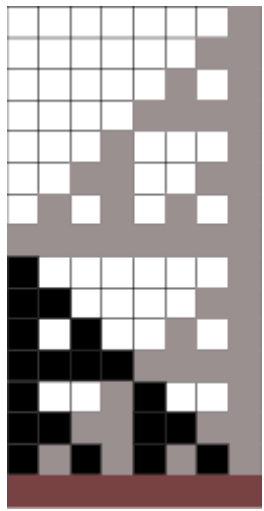

Fig.5. the Rule 102 based solution n=8

Similarly, we can also rewrite the transition function into this algebraic form:

$$\begin{cases} \delta\left(\begin{bmatrix}a\\b\end{bmatrix}\begin{bmatrix}c\\d\end{bmatrix}\begin{bmatrix}e\\f\end{bmatrix}\right) = \begin{bmatrix}e + c(\mathrm{mod}\ 2)\\d + b(\mathrm{mod}\ 2)\end{bmatrix} \\ \delta\left(\$\ \begin{bmatrix}c\\d\end{bmatrix}\begin{bmatrix}e\\f\end{bmatrix}\right) = \begin{bmatrix}c + e(\mathrm{mod}\ 2)\\c\end{bmatrix} \\ \delta\left(\begin{bmatrix}a\\b\end{bmatrix}\begin{bmatrix}c\\d\end{bmatrix}\$\right) = \begin{bmatrix}c\\b + d(\mathrm{mod}\ 2)\end{bmatrix} \end{cases}$$

And the rule 102 transition table can be got as Table 1.:

Table 1. The Rule 102 transition table

| Q | Right side | | | | | G | Right side | | | | | X₁ | Right side | | | |
|---|---|---|---|---|---|---|---|---|---|---|---|---|---|---|---|---|
| | | G | Q | X₁ | * | | | G | Q | X₁ | * | | | G | Q | X₁ | * |
| Left side | G | G | Q | | | Left side | G | Q | G | | G | Left side | G | F | X₁ | X₁ | |
| | Q | G | Q | | | | Q | Q | G | G | G | | Q | | | X₁ | |
| | X₁ | | X₁ | X₁ | | | X₁ | X₁ | | F | F | | X₁ | G | Q | Q | |
| | * | G | Q | | | | * | X₁ | | | | | * | F | X₁ | X₁ | |

5. The hybrid solution of Rule 60 and Rule 102

Based on the work of former text, we combine the Rule 60 and 102 to construct a solution for 2 generals that are at left and right of the line of $2^n$ FSSP respectively, we also have four states: the state of general (the gray color) ; the state of quiescent (the white color), the state of $X_1$ (the black color states), the state of firing(the dark red color), Here, we give a snapshot for synchronization operation for this algorithm on eight cells:

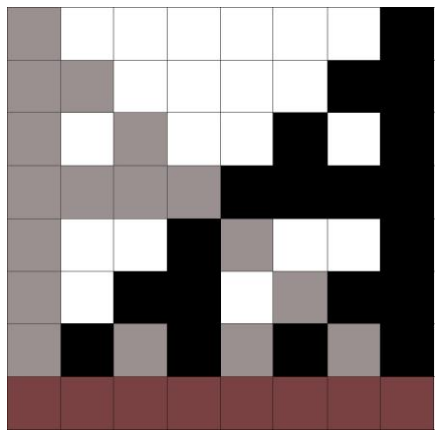

Fig.6. the Rule 102 and Rule 60 based solution n=8

We can draw the conclusion safely that this algorithm can short the synchronization time for 2-general, we define the state of general 1 is G, the state of general 2 is H, the state of quiescent is Q, the state of firing is F, the boundary state is *, the transition tables are as follows:

Table 2. The Rule 102 and Rule 60 transition tables

| H | Right Side | | | |
|---|---|---|---|---|
| | G | H | Q | * |
| Left Side G | F | G | | F |
| Left Side H | H | Q | | H |
| Left Side Q | H | Q | H | H |
| Left Side * | | | | |

| G | Right Side | | | |
|---|---|---|---|---|
| | G | H | Q | * |
| Left Side G | Q | H | Q | |
| Left Side H | G | F | G | |
| Left Side Q | | | G | |
| Left Side * | G | F | G | |

| Q | Right Side | | | |
|---|---|---|---|---|
| | G | H | Q | * |
| Left Side G | G | | G | |
| Left Side H | H | | H | |
| Left Side Q | | H | Q | |
| Left Side * | | | | |

In terms of the transition tables we can get the states change tables as follows:

Table 2. The Rule 102 and Rule 60 states transition tables

| G | Q | Q | Q | Q | Q | Q | H |
|---|---|---|---|---|---|---|---|
| G | G | Q | Q | Q | Q | H | H |
| G | Q | G | Q | Q | H | Q | H |
| G | G | G | G | H | H | H | H |
| G | Q | Q | H | G | Q | Q | H |
| G | G | H | H | G | G | H | H |
| G | H | G | H | G | H | G | H |
| F | F | F | F | F | F | F | F |

6. Conclusion

An existence or non-existence of four-state firing squad synchronization protocol has been a long-standing, famous open problem for a long time. In this paper, we have presented a 4-state solution to the firing squad which synchronizes the line of length $2^n$ in (n-1) step. Based on the result of the Rule 60 and rule 102 to FSSP, we combine with the Rule 60 and Rule 102 to give a solution for 2 generals that are at left and right of the line of $2^n$ FSSP respectively, in the end we also get states change tables. We think this is a very promising approach for the search of solutions designed for non minimal-time solutions with few states.